\documentclass[a4paper,fleqn,usenatbib]{mnras}
\usepackage{graphicx}
\usepackage{amssymb}
\usepackage{amsmath}
\usepackage[T1]{fontenc}
\usepackage{ae,aecompl}
\usepackage{natbib,times}
\usepackage{lscape}

\newcommand{\mc}{\multicolumn}

\begin{document}

\title[Radio luminosity function of BCGs]
{Radio luminosity function of brightest cluster galaxies}

\author[Yuan, Han, \& Wen]
{Z. S. Yuan$^{1,2}$, J. L. Han$^1$\thanks{E-mail: hjl@nao.cas.cn} 
and Z. L. Wen$^1$ 
\\
$^1$National Astronomical Observatories, Chinese Academy of Sciences, 20A
Datun Road, Chaoyang District, Beijing 100012, China\\
$^2$School of Astronomy, University of Chinese Academy of Sciences,
Beijing 100049, China
}

\date{Accepted 2016 May 10. Received 2016 May 9; in original form 2016 February 15}

\pagerange{\pageref{firstpage}--\pageref{lastpage}} \pubyear{2016}

\maketitle

\label{firstpage}


\begin{abstract}
  By cross-matching the currently largest optical catalog of galaxy
  clusters and the NVSS radio survey database, we obtain the largest
  complete sample of brightest cluster galaxies (BCGs) in the redshift
  range of $0.05<z\le0.45$, which have radio emission and redshift
  information. We confirm that more powerful radio BCGs tend to be
  these optically very bight galaxies located in more relaxed
  clusters. We derived the radio luminosity functions of BCGs from the
  largest complete sample of BCGs, and find that the functions depend
  on the optical luminosity of BCGs and the dynamical state of galaxy
  clusters. However, the radio luminosity function does not show
  significant evolution with redshift.
\end{abstract}


\section{Introduction}

Among many galaxies embedded in hot gas inside a galaxy cluster, the
brightest cluster galaxy (BCG) is the most massive and luminous galaxy
located near the center of the cluster \citep[e.g.][]{wh15a}. BCGs of
galaxy clusters differ from other elliptical galaxies in many aspects
because of unique cluster environments they inhabits in and the
evolution history they have experienced
\citep[e.g.][]{rmn+06,mmn13}. The nuclei of BCGs host supermassive
black holes \citep[e.g.][]{rmn+06}, so that BCGs manifest their active
nuclei by radio jets which is the feedback to cluster medium
\citep[see][]{mn07}. BCGs are more likely to be radio loud than other
galaxies \citep[e.g.][]{bwh81,bvk+07}.  The radio emission of BCGs is
related to both nuclei activities of BCGs and cluster properties.

\begin{table*}
\setlength{\tabcolsep}{1.5mm}
\begin{center}
\caption{A list of statistical studies of radio BCGs with a sample size over 100.}
\begin{tabular}{lccccc}
\hline
\mc{1}{l}{Authors} &\mc{1}{c}{Cluster sample} &\mc{1}{c}{Redshift range}  &\mc{1}{c}{Radio data} &\mc{1}{c}{Flux limit} &\mc{1}{c}{No. of radio BCGs}\\
\hline
Lin \& Mohr (2007)           &342 NORAS/REFLEX &$z<0.2$                &NVSS               &10 mJy         &122       \\
von der Linden et al. (2007) &625 C4 clusters  &$0.02<z<0.1$           &NVSS/FIRST         &5 mJy       & 252       \\
Best et al. (2007)           &484 C4 clusters  &$0.02<z<0.1$           &NVSS/FIRST         &5 mJy       & 252      \\
Croft et al. (2007)          &13,240 MaxBCG     &$0.1\le z\le0.3$       &FIRST              &               &2,615     \\
Antognini et al. (2012)      &13,823 MaxBCG     &$0.1\le z\le0.3$       &FIRST/NVSS         &0.75 mJy       &151      \\
Ma et al. (2013)             &685 X clusters   &$0.1\le z\le0.6$       &NVSS               &3 mJy          &357       \\
Hogan et al. (2015)          &720 REFLEX/(e)BCS&$0.03<z\le0.45$        &NVSS/SUMSS/ATCA/VLA&15 mJy         &437       \\[2mm]
This work                    &62,686 WH15       &$0.05<z\le0.45$        &NVSS/FIRST         &5 mJy          &7,138     \\
\hline
\end{tabular}
\end{center}
\label{tab1}
\end{table*}

Previously there has been much effort to decouple the effect of galaxy
properties and of cluster environment on radio emission of
BCGs. Without the large radio sky survey data, small samples of galaxy
clusters, usually less than a hundred, have been observed for
statistics in radio band \citep[e.g.][]{bwh81,zbo89,b90,bbl93}.
The fraction of BCGs being radio-loud above a given threshold of radio
flux density or luminosity has often been investigated for some
cluster samples to exam possible links between BCG radio emission and
the cluster cooling flows \citep[e.g.][]{pfe+98,mhr+09} or cluster
mass and X-ray luminosity \citep[e.g.][]{lm07,mmn+11,she+12} or
dynamic states \citep[e.g.][]{lm07,kvc+15}. \citet{mhr+09} found that
BCG radio luminosities were correlated with cluster cooling time, the
mass of supermassive black holes and also X-ray luminosity of strong
cool-core clusters. \citet{kvc+15} found that a larger fraction of
BCGs in relaxed clusters is radio loud than those in merging clusters.

Because of large data scatters, the statistics can be improved by
using large samples of BCGs observed by radio survey data.
\citet{vbk+07} and \citet{bvk+07} cross-identified 625 BCGs of the C4
cluster sample \citep{mnr+05} with the radio data of NVSS and FIRST
surveys \citep{ccg+98,bwh95}, and found that the fraction of radio
loud BCGs depends on optical luminosity or the stellar mass of BCGs,
but not on the cluster velocity dispersion.
\citet{abm12} got a sample of 151 FR-II type BCGs by looking at the
NVSS and FIRST images of the MaxBCG clusters \citep{kma+07}, and
concluded similarly that both the jet power and the radio loud
fraction is correlated with the $r$-band luminosity of BCGs but not
the cluster richness.
By cross-matching the MaxBCG cluster catalog \citep{kma+07} with the
FIRST radio survey data \citep{bwh95}, \citet{cdb07} obtained a very
large sample of 2,615 radio BCGs with $L_{\rm
  1.4GHz}>10^{23}$W~Hz$^{-1}$, and confirmed that the radio loud
fraction depends on the $r$-band absolute magnitude and hence the
converted stellar mass of BCGs, from about 5\% at
$10^{10.7}$M$_{\odot}$ to about 30\% at $10^{11.6}$M$_{\odot}$ and
that the fraction is larger for the BCGs in richer clusters.
\citet{mmn13} obtained a large sample of 357 radio BCGs by
cross-correlating galaxy clusters in eight X-ray catalogs and NVSS
radio sources, and confirmed that the radio fraction increases
moderately with redshift and cluster X-ray luminosity. They also found
that the radio power is greater in more massive clusters and at higher
redshifts, which implies possible redshift evolution of BCG radio
emission power.
Recently, \citet{heh+15} investigated 437 radio BCGs with
multi-frequency observational images, and found that the core emission
is more frequently associated with BCGs with [O III] emission that is
the canonical tracer for AGN activity. See Table~\ref{tab1} for a
summary of these previous samples.

More fundamental studies of the BCG radio emission or its difference
from other galaxies and their dependence on cluster properties should
work on the radio luminosity function, which is a measure of the
variation of BCG space density with radio luminosity. To construct a
good radio luminosity function a complete sample of objects is needed
with known redshifts as well as good measurements of radio emission.
Such studies have often been done for galaxies and AGNs
\citep[e.g.][]{con89,dbk09,bh12,sri+12,vfs+12,msn+12,bks+14}.  The
radio luminosity functions for AGNs and star-forming galaxies have
been found to be very different \citep[e.g.][]{sjc+02,ms07} but no
significant evolution with redshift
\citep[e.g.][]{pmk+11,mj11,mjb13,pbk+15,pmj+16}. The radio luminosity
functions for BCGs have been tried based on small samples of radio
measurements \citep[e.g.][]{bbl93,lm07}.

In this paper, we cross-match the currently largest optical catalog of
galaxy clusters \citep{wh15b}  with the largest radio survey
database of the NVSS and FIRST \citep{ccg+98,bwh95}. As shown in
Section 2, we get the largest complete sample of radio BCGs. With such
a sample, we study the possible dependence of BCG radio emission on
BCG properties and cluster environment in Section 3, and then work on
the radio luminosity function of BCGs in Section 4. The conclusions
are given in Section 5.

Throughout this paper, we assume a $\Lambda$CDM cosmology, taking
$H_0=100 h {\rm km~s}^{-1} {\rm Mpc}^{-1}$, with $h=0.7$,
$\Omega_{\rm m}=0.3$ and $\Omega_{\Lambda}=0.7$.


\section{BCG sample and radio emission power}

Based on the photometric data of the Sloan Digital Sky Survey (SDSS)
Data Release 8 \citep[DR8,][]{aaa11}, \citet{whl12} identified 132,684
galaxy clusters. Recently, \citet[][hereafter WH15]{wh15b} have
updated the parameters of these clusters with spectroscopic redshift
data in DR12 \citep[]{aaa15} and further identified 25,419 new
clusters. In total there are 158,103 galaxy clusters in the WH15
cluster catalog. This sample of galaxy clusters is complete up to
redshift $z\sim0.5$ for massive clusters of $M_{500} > 2\times
10^{14}$~M$_{\odot}$ \citep[see Figure 6 in][]{whl12}, which was
further verified by their redshift distribution in Figure 7 of
\citet{wh15b}. Here we take 62,686 galaxy clusters with a redshift
$z\le0.45$ and a richness $R_{L*}\ge12$ in the BOSS DR12 sky region of
9,376 square degrees as the parent sample in the following study. The
cluster parameters, such as richness $R_{L*}$ and redshift $z$
(including 56,340 spectroscopic and 6436 photometric redshifts), of
these clusters are directly taken from \citet{wh15b}. As shown
in \citet{wh15b}, the cluster richness $R_{L*}$ is a good measure of
optical mass proxies with a scatter of 0.17 dex. The $r$-band absolute
magnitude of BCGs is corrected for redshift evolution by using $
M_{\rm r}^{\rm e}=M_{\rm r}^{\rm SDSS}+Qz $, with $Q=1.16$ as done in
\citet{wh15b}. The BCG dominance, defined as the difference of
absolute magnitudes of the first and second BCGs, $M_{\rm r,2}-M_{\rm
  r,1}$, can statistically indicate the dynamic state of galaxy
clusters \citep[][]{wh13}.

\begin{figure}
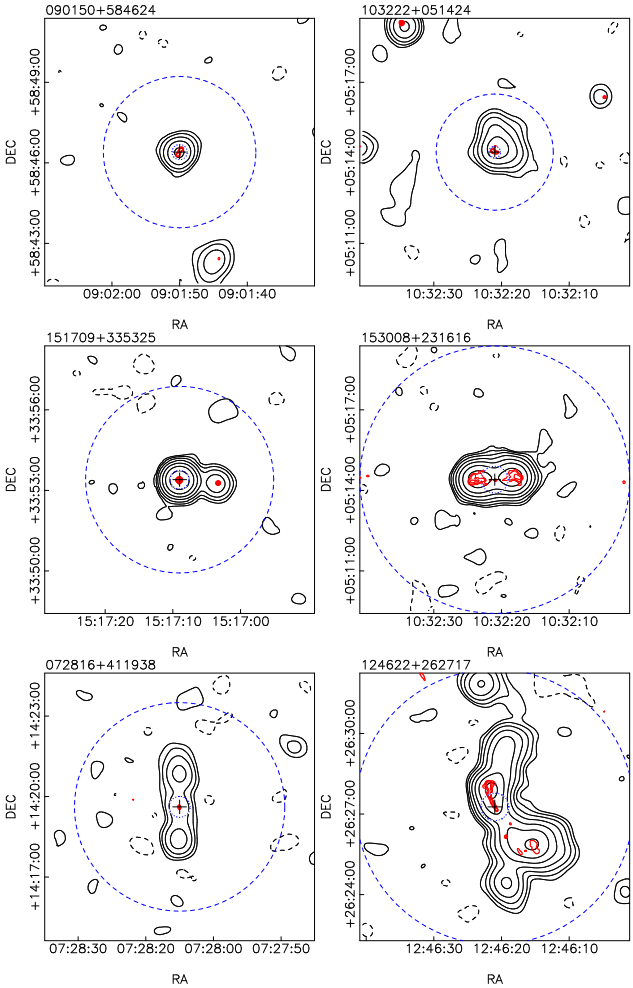

\centering
\includegraphics[angle=-90,width=0.23\textwidth]{fig1a.ps}
\includegraphics[angle=-90,width=0.23\textwidth]{fig1b.ps}
\includegraphics[angle=-90,width=0.23\textwidth]{fig1c.ps}
\includegraphics[angle=-90,width=0.23\textwidth]{fig1d.ps}
\includegraphics[angle=-90,width=0.23\textwidth]{fig1e.ps}
\includegraphics[angle=-90,width=0.23\textwidth]{fig1f.ps}
\caption{Example images for 6 radio BCGs. Low-resolution (black)
  contours are plotted for the NVSS data and high-resolution (red)
  contours for the FIRST data at levels of $\pm1,2,4,...$ mJy/beam,
  with the central cross indicating the optical position of a BCG.
  The big dashed circle indicates the projection distance of 500~kpc
  and the small dotted circle near the cross stands for 50~kpc from a
  BCG. Some radio sources are well coincident with BCGs within 50~kpc,
  and a small number of double or triple sources have the dominant
  flux from jets or lobes outside 50~kpc from a BCG.}
\label{fig1}
\end{figure}

To get the radio emission flux densities of these BCGs, we cross-match
the optical BCG sample with the NVSS radio source catalog which is
99\% complete above 3.5 mJy \citep{ccg+98}. Our approach is quite
similar to those in \citet{bkh+05b}, and we obtain our radio BCG
sample with a few steps. First, the radio flux density limit of 5 mJy
is adopted to ensure two individual components of a BCG above 2.5~mJy
not being missed, and we find the NVSS radio sources within a
projection distance of 500 kpc from BCGs, which in general is
sufficiently large to pick up radio emission components from
BCGs. 15,387 BCGs are found to be associated with 18,600 radio
sources. Naturally some of these sources come from BCGs, but others
are background radio sources or the radio emission from other galaxies
instead of BCGs. Among them, 12,885 BCGs have only one NVSS source,
and 1,985 BCGs have two and 517 BCGs have three or more NVSS sources
within the projected 500 kpc.

\begin{figure}
\centering
\includegraphics[angle=-90,width=0.40\textwidth]{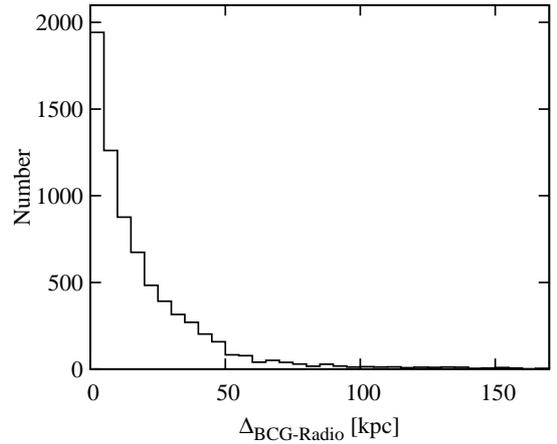}
\caption{Offsets between optical positions and radio coordinates
  for 7,138 radio BCGs.}
\label{fig2}
\end{figure}

 Second, we check the physical association more carefully, and adopt a
 more restrict criterion for the association which is the projection
 distance less than 50~kpc. For BCGs with only one NVSS source, 5,410
 BCGs have one coincident NVSS source within 50 kpc which are all
 accepted as radio BCGs; 6,526 BCGs have a NVSS source outside 100~kpc
 from the optical position are directly declined as radio BCGs; 257 of
 the rest 949 BCGs with one NVSS source between 50 and 100~kpc are
 accepted as radio BCGs, because either high-resolution FIRST images
 indicate the physical association or the NVSS source peak is
 coincident with an optical BCG within 50~kpc (see Figure~\ref{fig1}
 for example images).
 For 1,985 BCGs with two NVSS sources, we found that 839 of 1,985 BCGs
 have one NVSS source within 50~kpc though there is the other source
 beside with a very different flux density, which we adopt as radio
 BCGs with a flux of the coincident source; 28 BCGs are coincident
 with one of the double sources which have a similar flux densities
 (within a factor of 2) coming from radio lobes or jets of a radio
 galaxy at the middle position of two sources, and hence are not radio
 BCGs; 226 BCGs with two NVSS sources outside 50 kpc are adopted as
 radio BCGs, because the FIRST and/or NVSS images clearly indicate the
 association of a BCG and double sources.
 Similarly, for 517 BCGs with three or more than three NVSS sources in
 500 kpc, we adopt: 1) 275 of them as radio BCGs because one NVSS
 source is within 50 kpc from a BCG (excluding 5 BCGs coincident with
 one of double sources); 2) 91 BCGs with NVSS sources outside 50 kpc
 are adopted as radio BCGs because the FIRST and/or NVSS image clearly
 show the association of the BCGs and jets or core.
 In short, we identified 7,138 radio BCGs in total including 5,667
 BCGs with one NVSS source, 1,105 BCGs with two and 366 BCGs with
 three or more NVSS sources, see a list in Table~\ref{tab2} for the
 optical and radio parameters. The offset of radio sources from
 optical BCG positions are shown in Figure~\ref{fig2}.

\begin{table*}
\setlength{\tabcolsep}{1.5mm}
\begin{center}
\caption{The optical and radio parameters for 7,138 radio BCGs (see
  online Supporting Information for the full table).}
\begin{tabular}{rrrrrcccc}
\hline \mc{1}{c}{RA} &\mc{1}{c}{Dec} &\mc{1}{c}{$z$}
&\mc{1}{c}{$R_{L*}$} &\mc{1}{c}{$M\rm_{\rm r}^{\rm e}$}
&\mc{1}{c}{$M_{\rm~r,2}-M_{\rm~r,1}$} &\mc{1}{c}{$S_{\rm 1.4~GHz}$}
&\mc{1}{c}{$P_{\rm 1.4~GHz}$} &\mc{1}{c}{$\Delta_{\rm BCG -
    Radio}$}\\ \mc{1}{l}{(J2000)} &\mc{1}{c}{(J2000)} &\mc{1}{c}{ }
&\mc{1}{c}{(L*)} &\mc{1}{c}{(mag)} &\mc{1}{c}{(mag)} &\mc{1}{c}{(mJy)}
&\mc{1}{c}{($10^{24}\rm~W~Hz^{-1}$)} &\mc{1}{c}{(kpc)}\\
\hline
  0.20327&   8.66654& 0.4085&  31.24& -22.95& 0.35&   11.8&   7.25&  11.4\\
  0.20376&  -3.01915& 0.3732&  22.17& -22.78& 0.66&   15.6&   7.69&   5.7\\
  0.20552&  -0.84525& 0.4110&  28.58& -23.18& 1.00&   24.2&  15.09&  33.9\\
  0.27341&  34.46576& 0.2474&  23.53& -23.09& 1.44&   37.6&   7.00&  14.8\\
  0.31298&  -8.44618& 0.3288&  17.69& -23.01& 0.43&  151.1&  54.95&  19.0\\
  0.32537&  28.99514& 0.4247&  43.16& -23.92& 1.22&   10.9&   7.36&   5.0\\
  0.42376&   1.98004& 0.4377&  29.31& -23.17& 0.28&    5.5&   4.00&  69.8\\
  0.47782&   5.66544& 0.2399&  14.48& -22.89& 1.01&   29.3&   5.08&  42.5\\
  0.53473&  19.29001& 0.1469&  12.44& -23.81& 0.55&   17.6&   1.00&  37.3\\
  0.54585&  27.82813& 0.3354&  27.48& -23.37& 1.08&    5.4&   2.06&  16.3\\
  0.56685&  14.85623& 0.2967&  13.02& -23.10& 0.74&   37.6&  10.71& 114.6\\
  0.57405&  -7.26382& 0.3257&  19.03& -23.38& 0.88&   14.1&   5.01&  19.3\\
  0.60304&  -0.54798& 0.2902&  27.76& -23.04& 0.18&   19.0&   5.14&  24.4\\
  0.67632&  -0.22217& 0.2989&  26.52& -23.36& 0.82&   40.8&  11.83&  20.5\\
  0.68671&  -1.82153& 0.3941&  17.68& -23.05& 0.84&   14.2&   7.99&   6.9\\
\hline
\end{tabular}
\end{center}
\label{tab2}
\end{table*}

\begin{figure}
\centering
\includegraphics[angle=-90,width=0.45\textwidth]{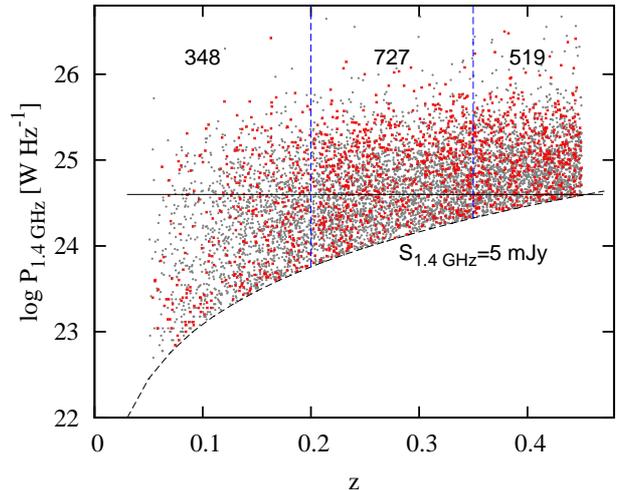}
\caption{Radio emission power of 7,138 BCGs. All BCGs have a radio
  flux density $S_{\rm 1.4~GHz}>5$~mJy. BCGs of a complete sample of
  1,594 massive clusters of $M_{500} > 2\times 10^{14}$~M$_{\odot}$
  (equivalent to $R_{L*}>35$) are indicated by (red)
  crosses. Sub-samples for three redshift ranges are divided by
  vertical dashed lines, which will be used for checking the redshift
  evolution of the BCG radio luminosity function in Section 4. Above
  a black line of $\log P_{\rm 1.4~GHz} =24.6$ is a power-limited
  complete sample of radio BCGs up to $z=0.45$.}
\label{fig3}
\end{figure}

We take flux densities of radio sources from the low resolution survey
NVSS to calculate radio emission power of BCGs. Note that the flux
densities of identified multiple radio components of a BCG have to be
added together to get $S_{\rm 1.4~GHz}$, and then the radio power is
obtained through
\begin{equation}
P_{\rm 1.4~GHz}=4\pi D_{\rm L}^{2}\times S_{\rm 1.4~GHz}\times (1+z)^{1-a},
\label{power}
\end{equation}
here $P_{\rm 1.4~GHz}$ is in W~Hz$^{-1}$, $D_{\rm L}=(1+z)
\frac{c}{H_0} \int_{0}^{z}\frac{dz'}
     {\sqrt{\Omega_{m}(1+z')^3+\Omega_{\Lambda}}}$ is the luminosity
     distance of a cluster at a redshift $z$, $S_{\rm 1.4~GHz}$ is the
     radio flux at 1.4 GHz from the NVSS, $(1+z)^{(1-a)}$ is the {\it
       k}-correction term with the spectral index $a$ of radio
     BCGs. We adopt the statistical mean of $a=0.74$ obtained by
     \citet{lm07} for all sources. The radio power distribution of
     7,138 BCGs is shown in Figure~\ref{fig3}.

Among them, 1,594 radio BCGs come from massive clusters of $M_{500} >
2 \times 10^{14}$~M$_{\odot}$, which form a radio-flux-limited
complete sample of BCGs that can be used for deriving the radio
luminosity function of BCGs in Section 4. We are confident that the
massive cluster sample is nearly 100\% complete during the cluster
identification \citep{whl12}, and the detection of radio emission for
$S_{\rm 1.4~GHz}>5$~mJy is at least 95\% complete in the
low-resolution NVSS survey even when BCGs have two components. On the
other hand, the radio detection of BCGs is power-limited complete
above a threshold of $\log P_{\rm 1.4~GHz} >24.6$ up to $z=0.45$.

\section{Fraction of radio BCGs and dependence on BCG and cluster properties}

As shown above, only 7,138 BCGs (11.4\%) of 62,686 galaxy clusters of
$z<0.45$ have radio emission with a flux density of $S_{\rm
  1.4~GHz}>5$~mJy. No doubt that more BCGs can be detected in radio
via more sensitive observations. In this section we check the possible
dependence of the radio fraction on BCG characteristics and cluster
properties. 

\begin{figure}
\centering
\includegraphics[angle=-90,width=0.45\textwidth]{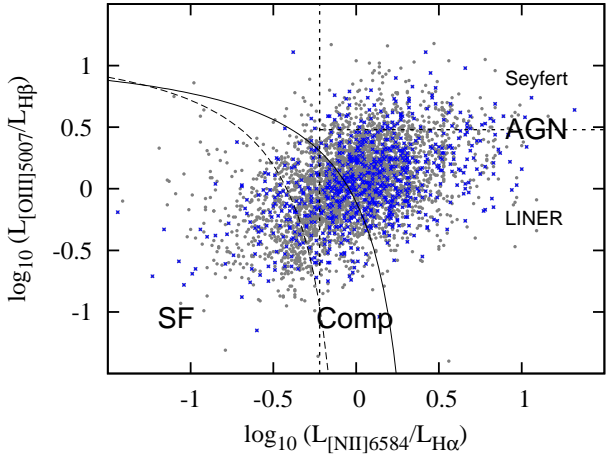}
\caption{The BPT diagram of emission line ratios for 2,763 BCGs which
  have a spectrum with a median signal to noise ratio greater than
  10. Among them 619 BCGs have their radio emission detected with
  $S_{\rm 1.4~GHz} > 5$~mJy, as indicated by little (blue)
  crosses. The solid curve is the demarcation for defining
  star-forming (SF) galaxies \citep{kds+01}, and the dash curve is the
  demarcation for classifying AGNs \citep{kht+03}. Dotted lines on
  $L_{\rm [O III]5007}/L_{\rm H\beta}=3$ and $L_{\rm [N
      II]6584}/L_{\rm H\alpha}=0.6$ are used to dividing LINERs and
  Seyferts.}
\label{fig4}
\end{figure}

\subsection{Radio fraction of spectroscopically classified BCGs}

Most BCGs are elliptical galaxies with active nuclei, showing AGN
properties. Nevertheless a few percent of galaxy clusters show a
cooling flow near the center of clusters \citep[see][]{f94,aem11},
which is probably related to the on-going or post star formation
features of BCGs \citep[e.g.][]{oqo+10,lwhm12,lmm12}. It is intriguing
to know how many BCGs have AGNs and how many of them possess star
formation features.

\begin{table}
\setlength{\tabcolsep}{1.5mm}
\caption{Radio fraction of spectroscopically classified BCGs}
\label{tab3}
\begin{center}
\begin{tabular}{lrrrr}
\hline
\hline
\mc{1}{l}{BCG samples} &\mc{1}{c}{SF} &\mc{1}{c}{Comp} &\mc{1}{c}{AGN} &\mc{1}{c}{sum}\\
\hline
BCGs with 4 lines detected                    &357   &1103  &1303  &2763 \\[1mm]
No. of Radio BCGs                                &67    &161   &391   &619 \\
Radio fraction                            &18.8\%&14.6\%&30.0\%&22.4\%\\[1mm]
BCG No. of $\log P_{\rm 1.4 GHz}>24.6$          &25    &57    &151   &233 \\
Radio fraction                            &7.0\%&5.2\% &11.6\%&8.4\%\\
\hline
\end{tabular}
\end{center}
\end{table}

In general galaxies can be broadly classified as star-forming
galaxies, radio-loud AGNs, and composites
\citep[e.g.,][]{mg00,sjc+02,bkh+05b,ms07} according to spectra of
galaxies or their nuclei. The BPT diagram of line ratios \citep{bpt81}
has widely been used as a diagnostic to separate radio-loud AGNs from
star-forming galaxies \citep[e.g.,][]{kht+03,bvk+07,bh12}. In our
parent sample of 62,686 galaxy clusters, 56,340 BCGs have their
spectra observed already. However, to get the line ratios among
H$\alpha$, H$\beta$, [OIII]5007 and [NII]6584, the spectra of BCGs
should have a good signal-to-noise ratio. From the value-added
spectroscopic
catalogs\footnote{http://wwwmpa.mpa-garching.mpg.de/SDSS/DR7/}
produced by a research group from the Max Planck Institute for
Astrophysics and the Johns Hopkins University
\citep[see][]{thk+04,bcw+04}, 2,763 BCGs have a spectrum with a median
signal-to-noise ratio per pixel of the whole spectrum $sn_{\rm
  median}>10$ and with the pipeline warning flags $zWarning=0$, from
which four spectral lines of H$\alpha$, H$\beta$, [O III]5007 and [N
  II]6584 are significantly detected. The BPT diagram of 2,763 BCGs is
plotted in Figure~\ref{fig4}. According to the demarcation for
defining star-forming galaxies and AGNs \citep{kds+01,kht+03}, 1,303
(47.2\% of 2,763) BCGs have a AGN, 357 BCGs (12.9\%) have star-forming
features, and the rest 1,103 BCGs are composite galaxies. Among all
these BCGs, 619 (22.4\% of 2,763) BCGs have radio emission detected
above $S_{\rm 1.4~GHz} = 5$~mJy, as indicated in Figure~\ref{fig4}.

The subsample of BCGs with the four lines consists of bright nearby
(majority $z<0.2$) BCGs, not a complete sample by any
means. Nevertheless, the statistics on the radio fraction of these
BCGs given in Table~\ref{tab3} and the BPT diagram in
Figure~\ref{fig4} can at least tell that not all radio BCGs possess
AGNs, and that 10.8\% (i.e. 67/619) radio BCGs are star-forming
galaxies, not necessary just elliptical galaxies. The AGN percentage
is larger for radio BCGs (i.e. 63.2\% = 391/619) than for BCGs
in general (i.e. 47.2\% = 1303/2763).

\begin{figure*}
 \begin{center}
 \includegraphics[angle=-90,width=0.78\textwidth]{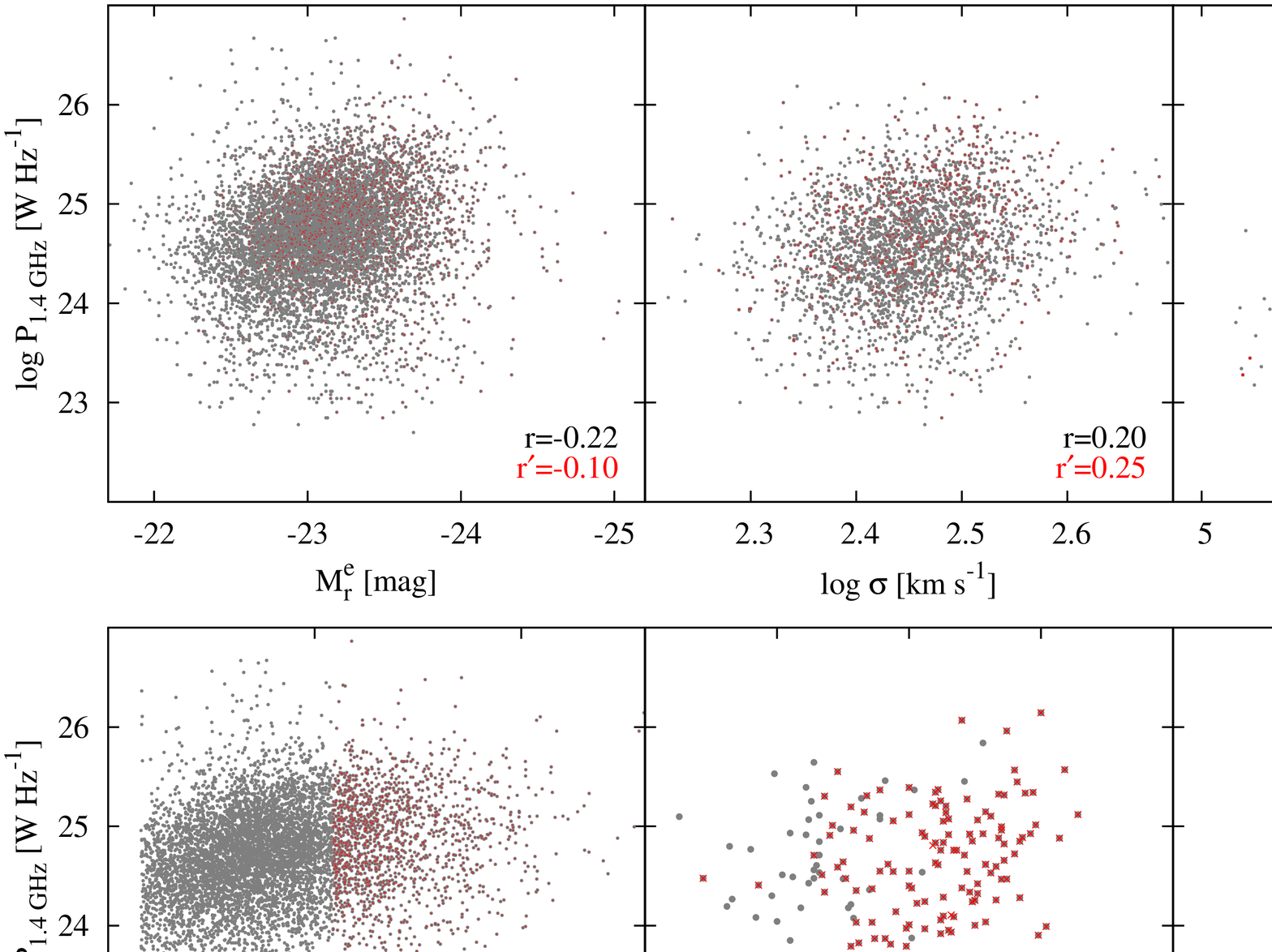}
 \caption{Radio power of BCGs are plotted against BCG optical
   parameters in {\it top panels} ({\it left} for the evolution
   corrected $r$-band absolute magnitude ${\rm M_r^e}$ , {\it middle}
   for stellar velocity dispersion $\sigma$, and {\it right} for the
   luminosity of [O III] lines $\rm L_{[O~III]}$), cluster properties
   in {\it middle rank panels} ({\it left} for richness $R_{L*}$, {\it
     middle} for X-ray estimated mass $\rm M_{\rm X}$ and {\it right}
   for SZ-estimated mass $\rm M_{\rm SZ}$) and cluster dynamic states
   in {\it bottom panels} ({\it left} for BCG dominance $\rm
   M_{r,2}-M_{r,1}$, {\it middle} for optical dynamical parameter
   $\Gamma$ from \citet{wh13}, and {\it right} for the offset $\Delta$
   of X-ray peak from BCG), if these parameters are available (see
   text). The radio BCGs in a complete sample of massive clusters of
   $M_{500} > 2 \times 10^{14}$~M$_{\odot}$ are indicated by (little
   red) crosses.  The Spearman correlation coefficients are given in
   the bottom-right corner of each panel for both all radio BCGs ($r$)
   and the radio-flux-limited complete sample ($r'$). }
 \label{fig5}
\includegraphics[angle=-90,width=0.78\textwidth]{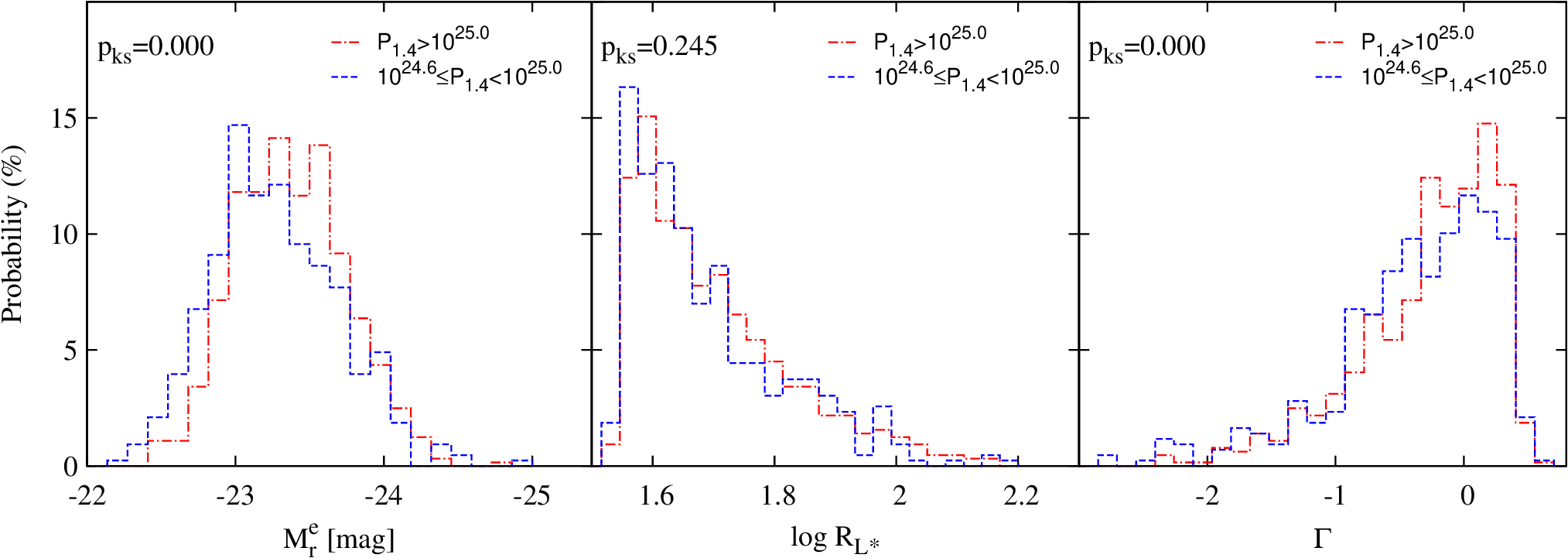}
\caption{The probability distributions for two power-limited complete
  sub-samples of a massive cluster sample of $M_{500} > 2 \times
  10^{14}$~M$_{\odot}$ along the evolution corrected $r$-band absolute
  magnitude ${\rm M_r^e}$ ({\it left}), cluster richness $R_{L*}$
  ({\it middle}) and the cluster dynamic state parameter $\Gamma$
  ({\it right}). The KS-test probability for the two subsamples is
  given in the upper-left corner of each panel.}
\label{fig6}
\end{center}
\end{figure*}

\subsection{Radio power versus optical characteristics of BCGs}

The BCGs are unusual galaxies.  By using the large sample of 7,138
radio BCGs, we here investigate the possible dependence of BCG radio
power on the absolute magnitude, velocity dispersion, and luminosity
of [O III]5007 line. The absolute $r$-band magnitude is an indicator
of stellar mass of galaxies and is also related to the mass of a
central black hole \citep[e.g. see Eq.(8-10) in][]{mhr+09}. The
stellar velocity dispersion has been well-established to be related to
the mass of central black hole \citep{tgb+02,mh03}. While the high
ionization [O III] forbidden line is emitted by filamentary gas
related to jets \citep{m93}, hence it is the indicator of AGN activity
\citep[e.g.][]{mr95}. Beside the evolution corrected $r$-band absolute
magnitude of 7,138 radio BCGs, we obtain the stellar velocity
dispersion $\sigma$ of 2,884 BCGs and the luminosity of [O III]5007
line of 1,237 BCGs from the value-added spectroscopic catalogs
pproduced by a research group from the Max Planck Institute for
Astrophysics and the Johns Hopkins University
\citep[see][]{thk+04,bcw+04}.

As shown the top panels of Figure~\ref{fig5}, the radio power data in
all three panels are rather scattered against the absolute magnitude,
the velocity dispersion and [O III]5007 line luminosity of BCGs. Our
results are consistent with the plots previously obtained for a small
sample of BCGs by \citet{mhr+09} and large samples of BCGs by
\citet{cdb07}, \citet{abm12} and \citet{heh+15}. The mild or strong
correlation between radio power and the absolute magnitude found for
radio galaxies \citep[e.g.][]{cff89} is not shown for radio BCGs, even
for the complete sample of radio BCGs in massive clusters. No
significant correlation between the radio power and stellar velocity
dispersion is found for BCGs. The Spearman rank-order correlation
coefficient $r$ in each panel tells how strong the correlation is.
Nevertheless, a weak correlation ($r'=0.41$) exists between the
radio power and luminosity of [O III] line of BCGs as shown in the top
right panel of Figure~\ref{fig5}, though not as strong as for radio
galaxies \citep{sbr+89,m93,wrb+99,ssk+13}. Very probably line emission
is fundamentally related with core radio emission \citep{heh+15} of
high-excitation radio galaxies \citep{bh12}. The radio power of BCGs
in this paper includes all radio emission components, not just limited
to core.

Note that the fraction of radio loudness of BCGs has been related to
the absolute magnitude or stellar mass
\citep{bvk+07,mhr+09,kvc+15}. We will check it together with cluster
richness in Sect.~\ref{rc}, and here show the probability distribution
of two power-limited complete samples of radio BCGs along the absolute
magnitude in the left panel of Figure~\ref{fig6}. The KS test tells
that the BCG samples of $24.6<\log P_{\rm 1.4~GHz} <25.0$ and of $\log
P_{\rm 1.4~GHz} >25.0$ have different distributions of the absolute
magnitude or stellar mass with an obvious shift of more powerful radio
BCGs to optical brighter magnitude, which echos with the previous
conclusion \citep[e.g.,][]{bvk+07,vbk+07,abm12,heh+15} that the radio
loud fraction is larger for more massive BCGs for a given flux density
threshold.

\subsection{Radio power of BCGs versus cluster properties}

We investigate here the possible correlation between the radio power
of BCGs and cluster environments. Most of previous studies have worked
on the X-ray cluster samples \citep[e.g.][]{lm07,mmn13,heh+15} and
check if the BCG radio emission or the fraction of radio BCGs depends
on cluster dynamical state or cluster mass. Observations of small
sample of clusters show that BCGs in relax clusters have higher
possibility to be radio loud than those in unrelaxed clusters and that
radio power of BCGs seems to correlate with dynamical parameters of
clusters \citep[e.g.][]{pfe+98,mhr+09, kvc+15}. Here we work on the
optical cluster sample.

Two classes of cluster parameters are investigated for their relation
with BCG radio emission power. First is on cluster mass. Mass of
galaxy clusters is in general estimated from the X-ray data
\citep[e.g.][]{rb02}. Up to now only a few thousand galaxy clusters
have their mass so estimated \citep[e.g.][]{vbe+09,mae+10,pap+11}. We
get X-ray estimated mass for 198 host clusters with radio BCGs from
the compiled catalog by \citet{wh15b} and the SZ-estimated masses for
90 clusters from the recent Planck cluster catalog \citep{paa+15}. The
cluster richness $R_{L*}$ derived from the total optical luminosity of
member galaxies is used as an optical mass proxy of clusters as
verified by \citet{wh15b}.

In the middle-rank of Figure~\ref{fig5}, the BCG radio power is
plotted against $R_{L*}$ and the cluster mass estimated from X-ray and
SZ observations. Data scattered but show weak correlations in all
three panels which indicates the tendency that more powerful radio
BCGs are hosted by more massive clusters. The BCGs in clusters with
smaller richnesses or less masses have a distribution of radio power
peaked at a smaller power than those in richer clusters, which is
consistent the results of \citet{lm07} and \citet{mmn13}. This is
understandable, because BCGs are brighter in richer clusters
\citep{whl12} which could produce a bit stronger radio emission (see
discussion above). It is intriguing, however, to see no correlation
between the radio power and cluster richness for the radio BCGs of a
complete sample of massive clusters (the left panel of middle-rank of
Figure~\ref{fig5}), and no difference is found for the cluster
richness distributions of two radio power-limited complete subsamples
of BCGs (see the middle panel of Figure~\ref{fig6}). Such a result
implies for no dependence of BCG radio emission on cluster richness
\citep{abm12}. Noticed that these two complete samples of BCGs are
hosted by clusters with only a small range of richness and that the
correlation is shown between radio power and $\rm M_{SZ}$. Therefore
whether there is any dependence of BCG radio power on richness or
cluster mass has to be concluded by further investigations of a large
cluster sample with a much large range of mass or richness.

Next is on dynamical states of galaxy clusters. In principle the
three-dimensional mass and velocity distributions of gas and member
galaxies should be measured to describe if a galaxy cluster is
dynamically relaxed or in a state of merging or dynamically
disturbed. In practice, one-dimensional measurements of velocity
distribution of member galaxies \citep[e.g.][]{ds88} and the
two-dimensional distribution of hot gas \citep[e.g.][]{me12} or member
galaxies \citep[e.g.][]{wh13} have been quantitatively analyzed for
cluster dynamical states. We here get the optical dynamical parameters
$\Gamma$ of 1,594 clusters from \citet{wh13} and from new calculations
and the offsets of BCGs from the peak of X-ray images for 149 host
clusters from \citet{pap+11}. We noticed that the BCG dominance $\rm
M_{r,2}-M_{r,1}$ can statistically be used as being an indicator of
dynamical parameter for galaxy clusters \citep[see][]{wh13}. These
dynamical parameters are not significant correlated with radio power
of BCGs, as shown in the lower panels of Figure~\ref{fig5}. However,
the distributions of optical dynamical parameters $\Gamma$ are
slightly different for the two radio-power-limited complete samples of
BCGs in massive clusters, as shown in the right panel of
Figure~\ref{fig6}. Slightly more radio-powerful BCGs are detected in
more dynamically relaxed clusters, which is consistent with the
results of \citet{kvc+15}.

\begin{figure}
\centering
\includegraphics[angle=-90,width=0.40\textwidth]{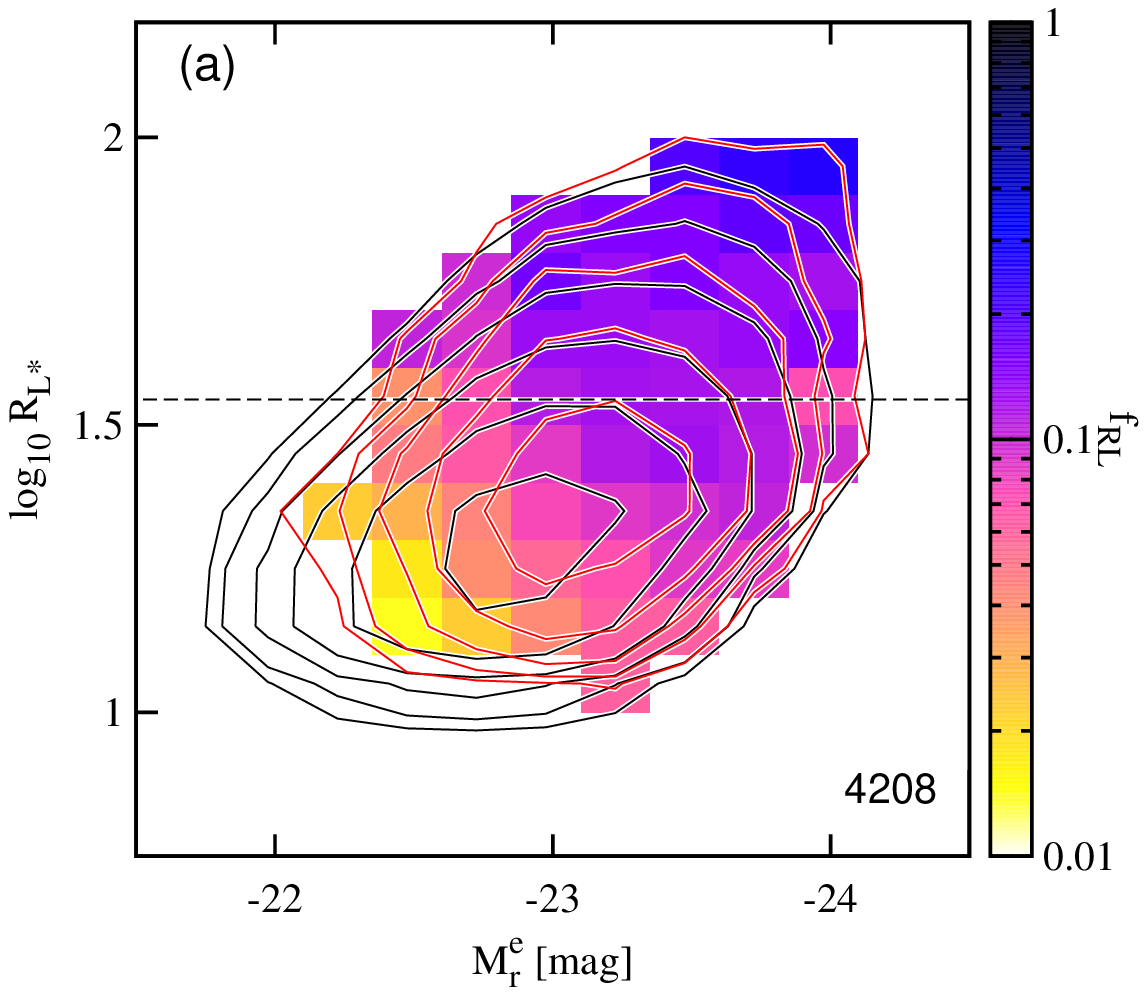}
\includegraphics[angle=-90,width=0.40\textwidth]{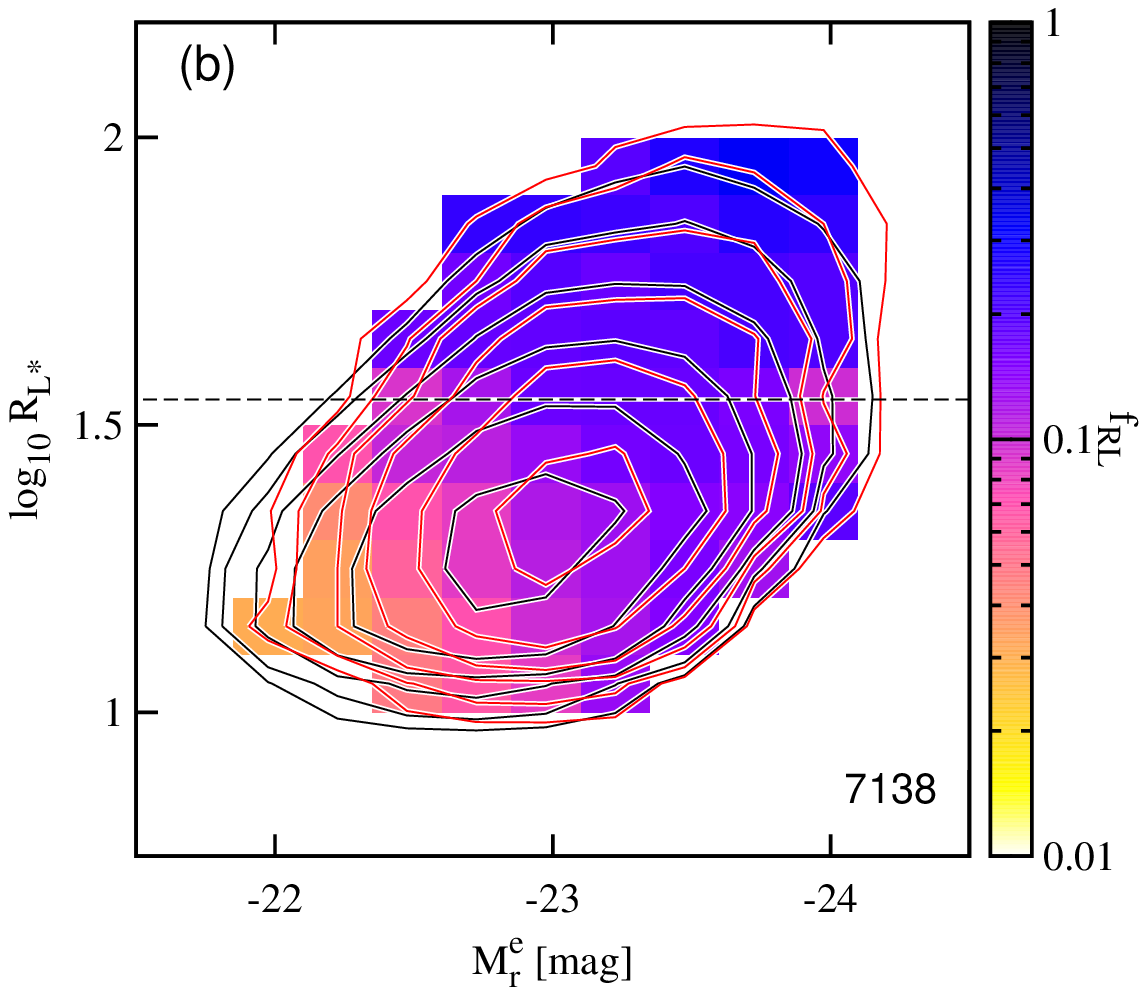}
\caption{The fraction of radio loud BCGs, $f_{\rm RL}$, varies with
  the $r$-band absolute magnitude of BCG and cluster richness. The
  {\it upper panel} is calculated by using 4,208 radio BCGs with a
  power-limited threshold of $P_{1.4\rm~GHz}>10^{24.6}\rm~W~Hz^{-1}$,
  and the {\it lower panel} by using 7,138 radio BCGs with a flux
  limit of $S_{1.4\rm~GHz}>5 \rm~mJy$, both of which are compared with
  62,686 BCGs in the optical parent cluster sample. Above the dash
  line of $R_{L*}=35$ is a complete BCG sample of massive clusters of
  $z<0.45$. A pixel is shown only if there have more than 10 radio
  BCGs. Overlaid are red contours for the number distribution of radio
  BCGs at levels of $5\times2^n$ (n =0,1,2,3, 4, 5) per pixel, and
  black contours for 62,686 BCGs of the parent sample at levels of
  $5\times2^n \times D$ (n =0,1,2,3,4,5), and $D=62,686/4,208$ or
  $62,686/7,138$ for the two samples, respectively. }
\label{fig7}
\end{figure}

\subsection{Fraction of radio loud BCGs versus BCG magnitude and cluster richness}
\label{rc}

Because of large data scatter, it is very hard to correlate the radio
power with BCG properties and cluster environments as shown above. The
fraction of radio loudness, $f_{\rm RL}$, defined as the percentage or
the ratio between the radio detected objects and the full sample, has
been used to check the possible dependence of radio emission on other
galaxy properties \citep[e.g.][]{bkh+05a}. The fraction of radio BCGs
has been related to either the absolute magnitude of BCGs
\citep[e.g.][]{vbk+07,bvk+07,cdb07,abm12} or the X-ray luminosity and
richness of galaxy clusters \citep[e.g.,][]{mmn13}, but not both
yet. Here we take this large sample of radio BCGs to study the
dependence of radio laud fraction in two-dimensions on both BCG and
cluster properties.

Remember that 7,138 radio loud BCGs were identified from a parent
sample of 62,686 optical clusters by using the NVSS and FIRST survey
data with a total flux density threshold of $S_{1.4\rm~GHz}>5
\rm~mJy$, and 4,208 of which are above the radio power threshold of
$P_{1.4\rm~GHz}=10^{24.6}\rm~W~Hz^{-1}$ in the redshift range
$0.05<z\le0.45$ (see Figure~\ref{fig3}). We define the radio loud
fraction $f_{\rm RL}\,(\rm M_r^e,\, R_{L*})$ as the number ratio of
these radio BCGs to the BCG numbers of the parent cluster sample for a
given small range of the BCG magnitude and cluster richness. The radio
fraction is then checked in two-dimensions for its possible dependence
on the BCG luminosity and cluster richness. As shown in
Figure~\ref{fig7}, for both the power-limited radio BCG sample or the
flux-limited radio BCG sample, there is a very clear tendency that the
radio fraction increases with both BCG magnitude and cluster richness,
from $\sim$1\% in the lower-left corner to $\sim$20\% in the top-right
corner that is for very bright BCGs in massive rich clusters. To
disentangle their effects, the dependence of radio fraction on one
parameter should be checked in Figure~\ref{fig7} only in a small range
of the other parameter. Integrating data in Figure~\ref{fig7} over one
dimension give a global dependence of the radio fraction on the other
dimension, echoing the results previously obtained by
\citet{cdb07,vbk+07,bvk+07,abm12,mmn13}.

We conclude that the BCG mass and cluster environments only
statistically affect the BCG radio emission, but not through individual
cases.

\begin{figure*}
\begin{center}
\includegraphics[angle=-90,width=0.8\textwidth]{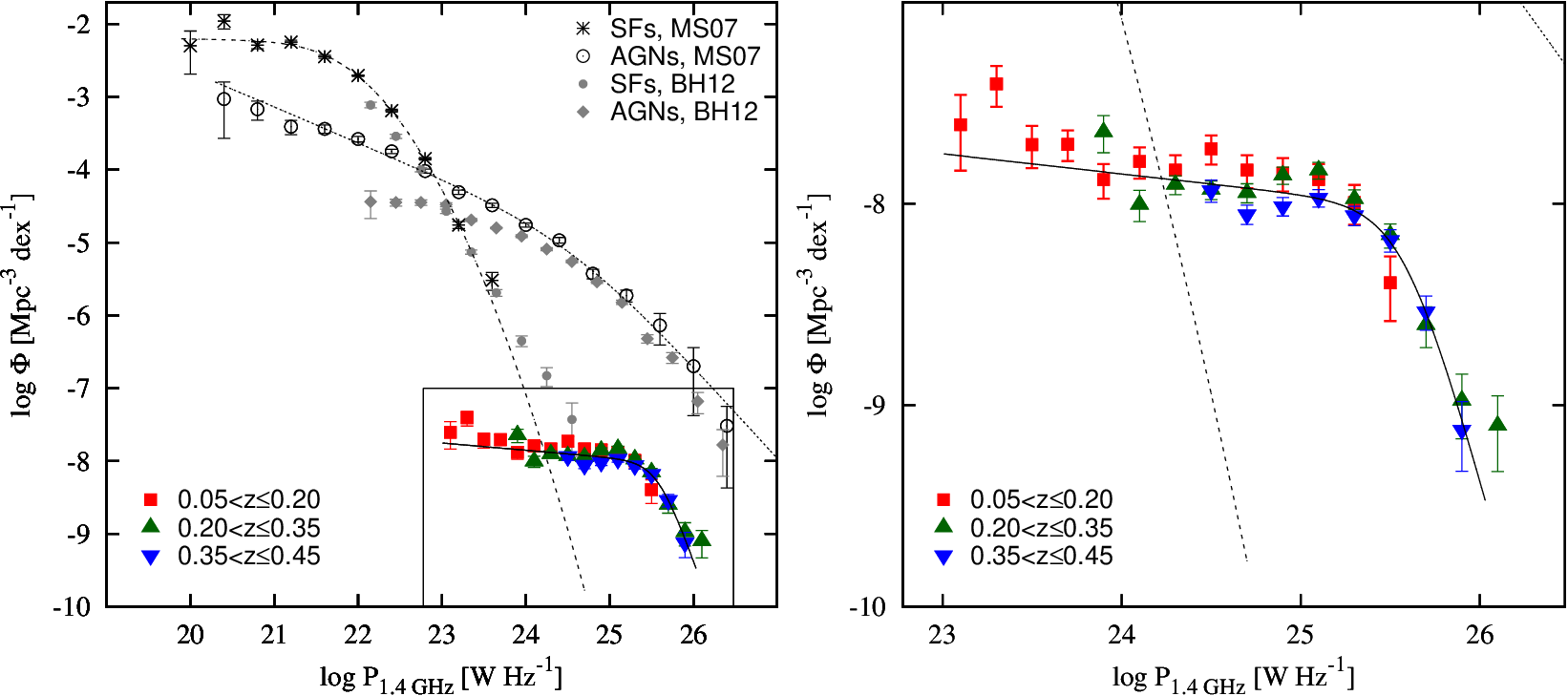}
\caption{Radio luminosity functions of BCGs derived for subsamples in
  three redshift ranges, compared with radio luminosity functions of
  star-forming galaxies and AGNs obtained by \citet{ms07} and
  \citet{bh12}. The solid line is the best fitting to radio luminosity
  functions of all BCGs, and dash and dotted lines stand for the
  fitting to the function for AGNs and star-forming in
  \citet{ms07}. An enlarged part for radio luminosity functions of
  BCGs is shown in the right panel. }
\label{fig8}
\includegraphics[angle=-90,width=0.8\textwidth]{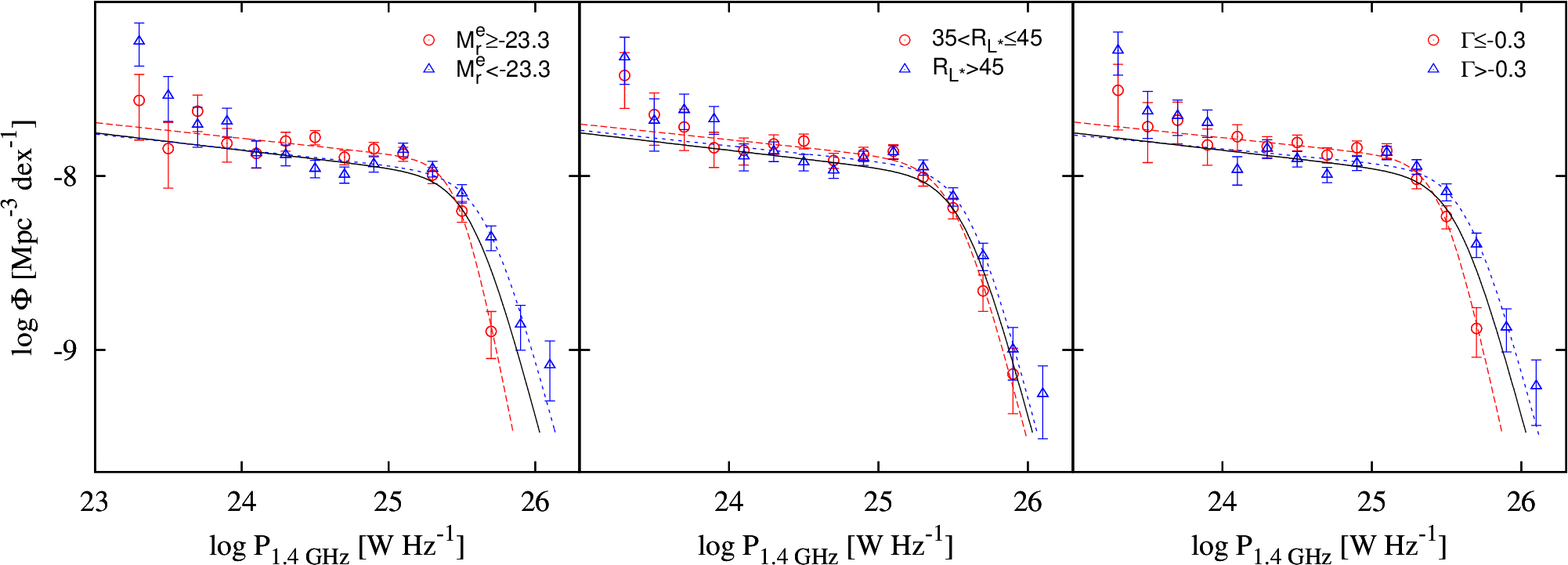}
\caption{Radio luminosity functions of BCGs in two subsamples with
  different ranges of BCG luminosity ({\it left}), cluster richness
  ({\it middle}) and dynamic states ({\it right}). The black solid
  lines is the best fitting for all radio BCGs, the same as that in
  Figure~\ref{fig8}, and dash and dot lines are best fittings for two
  subsamples. For better comparison, the radio luminosity functions of
  BCGs in two subsamples have been re-scaled to that of the full
  sample of BCGs according to source numbers.}
\label{fig9}
\end{center}
\end{figure*}

\begin{table*}
  \caption{Radio luminosity functions at 1.4 GHz for radio loud BCGs in three redshift ranges (only for more than 5 objects each bin).}
  \label{tab4}
\begin{center}
\begin{tabular}{ccccccc}
\hline
\mc{1}{l}{ }  &\mc{2}{c}{$0.05<z\le0.20$} &\mc{2}{c}{$0.20<z\le0.35$} &\mc{2}{c}{$0.35<z\le0.45$}\\
\mc{1}{c}{$\log P_{\rm 1.4~GHz}$} &\mc{1}{c}{$N$} &\mc{1}{c}{$\log \Phi$} &\mc{1}{c}{$N$} &\mc{1}{c}{$\log \Phi$} &\mc{1}{c}{$N$} &\mc{1}{c}{$\log \Phi$}\\
\mc{1}{c}{$\rm W~Hz^{-1}$} &\mc{1}{c}{ } &\mc{1}{c}{$\rm Mpc^{-3}~dex^{-1}$} &\mc{1}{c}{ } &\mc{1}{c}{$\rm Mpc^{-3}~dex^{-1}$} &\mc{1}{c}{ } &\mc{1}{c}{$\rm Mpc^{-3}~dex^{-1}$}\\
\hline
23.1   &10  &$-7.39_{-0.17}^{+0.12}$ &    &                     &     &                    \\[1mm]
23.3   &22  &$-7.34_{-0.10}^{+0.08}$ &    &                     &     &                    \\[1mm]
23.5   &22  &$-7.62_{-0.10}^{+0.08}$ &    &                     &     &                    \\[1mm]
23.7   &39  &$-7.63_{-0.08}^{+0.07}$ &    &                     &     &                    \\[1mm]
23.9   &28  &$-7.85_{-0.09}^{+0.08}$ &21  &$-7.69_{-0.11}^{+0.09}$ &     &                    \\[1mm]
24.1   &33  &$-7.78_{-0.08}^{+0.07}$ &34  &$-7.98_{-0.08}^{+0.07}$ &     &                    \\[1mm]
24.3   &26  &$-7.88_{-0.10}^{+0.08}$ &92  &$-7.88_{-0.05}^{+0.04}$ &     &                    \\[1mm]
24.5   &38  &$-7.72_{-0.08}^{+0.07}$ &86  &$-7.95_{-0.05}^{+0.04}$ &61   &$-7.95_{-0.06}^{+0.05}$ \\[1mm]
24.7   &31  &$-7.81_{-0.09}^{+0.07}$ &93  &$-7.91_{-0.05}^{+0.04}$ &77   &$-8.08_{-0.05}^{+0.05}$ \\[1mm]
24.9   &32  &$-7.79_{-0.08}^{+0.07}$ &105 &$-7.86_{-0.04}^{+0.04}$ &92   &$-8.00_{-0.05}^{+0.04}$ \\[1mm]
25.1   &28  &$-7.85_{-0.10}^{+0.08}$ &117 &$-7.81_{-0.04}^{+0.04}$ &100  &$-7.97_{-0.05}^{+0.04}$ \\[1mm]
25.3   &21  &$-7.97_{-0.11}^{+0.09}$ &87  &$-7.94_{-0.05}^{+0.04}$ &82   &$-8.05_{-0.05}^{+0.05}$ \\[1mm]
25.5   &7   &$-8.45_{-0.21}^{+0.14}$ &55  &$-8.14_{-0.06}^{+0.06}$ &60   &$-8.19_{-0.06}^{+0.05}$ \\[1mm]
25.7   &7   &$-8.45_{-0.21}^{+0.14}$ &21  &$-8.56_{-0.11}^{+0.09}$ &27   &$-8.54_{-0.09}^{+0.08}$ \\[1mm]
25.9   &    &                     &8   &$-8.98_{-0.19}^{+0.13}$ &6    &$-9.19_{-0.23}^{+0.15}$ \\[1mm]
26.1   &    &                     &6   &$-9.10_{-0.23}^{+0.15}$ &     &                    \\
Total  &348 &                     &727 &                    &519  &                     \\
\hline
\end{tabular}
\end{center}
\end{table*}

\section{Radio luminosity function of BCGs}

Noticed that the fraction of radio BCGs depends on the threshold of
radio observation. The radio power data are very scattered against BCG
and cluster properties. Luminosity function is an important tool to
study the evolution of space population. We now work on radio
luminosity function of BCGs, and check the dependence of the function
on the BCG and cluster properties.

The luminosity function $\Phi(P)$ stands for the comoving space
density of a kind of objects in a complete sample for a given
luminosity $P$ \citep[e.g.][]{ape+77,con89}. Considering the possible
cosmological evolution, the global average space density should be
calculated at the present epoch for the {\it local} luminosity
function \citep[e.g.][]{con89,ms07}.
As shown in section 2, we have got a complete sample of massive
clusters of $M_{500} > 2\times 10^{14}$~M$_{\odot}$ (i.e. $R_{\rm
  L*}>35$) within the redshift range of $0.05<z\le0.45$, from which
1,594 BCGs have been detected in radio surveys above the flux-limit of
$S_{1.4\rm~GHz}>5 \rm~mJy$ (see the crosses in Figure~\ref{fig3}).
This forms a complete radio BCG sample for radio luminosity function
of BCGs.

The radio luminosity functions $\Phi(P)$ are calculated in the
standard way, as $\Phi(P) = \sum_i 1/V_i$. Here $V_i$ is the volume in
which the $i$th BCG with a radio power between $P$ to $P+dP$ could be
detected \citep{sch68,con89,bh12}. For $N$ sources of a complete
sample detected in the redshift range of $z_{min} <z_i< z_{max}$ in a
given sky region, all of them have $V_i = V_{z_{max}}-V_{z_{min}}$, so
that $\Phi(P) = \sum^N_{i=1} 1/V_i = N /
(V_{z_{max}}-V_{z_{min}})$. Note here that the sky area of 9,376
square degree (i.e 2.85 sr) for the the complete sample of massive
clusters and hence the radio BCGs has to taken into account for
calculation of $V_i$.
To study if the radio luminosity functions of BCGs evolve with
redshift, we divide the complete sample into three sub-samples with
redshift ranges of $0.05<z\le0.20$, $0.20<z\le0.35$, and
$0.35<z\le0.45$. Their radio luminosity functions are listed in
Table~\ref{tab4} and plotted in Figure~\ref{fig8}. The uncertainties
of $\Phi(P)$ here include only the statistical Poissonian errors, and
hence are underestimated for some bins with small number of objects.

As shown clearly in Figure~\ref{fig8}, no evolution with redshift can
be found from radio luminosity functions obtained from the subsamples
of three different redshift ranges, which is consistent with previous
results for radio galaxies
\citep[e.g.,][]{scm+07,dbk09,mj11,sri+12}. Following \citet{ms07}, we
fit the radio luminosity functions of three subsamples together with a
two power-law analogous:
\begin{equation}
\Phi(P_{1.4~\rm GHz})=\frac{C_{0}}{(P_{1.4~\rm
    GHz}/P_{0})^{\alpha}+(P_{1.4~\rm GHz}/P_{0})^{\beta}},
\label{rlf}
\end{equation}
and obtained the best fitted parameters as
\begin{equation}
\begin{split}
C_{0}&=(9.4\pm1.4)\times10^{-9}~\rm Mpc^{-3}~dex^{-1}\nonumber;\\
P_{0}&=(39.9\pm5.5)\times10^{24}~\rm W~Hz^{-1};\\
\alpha&=3.43\pm0.79;\\
\beta&=0.12\pm0.06.
\label{parafit}
\end{split}
\end{equation}
Comparing to the radio luminosity functions of star forming galaxies
and AGNs obtained by \citet{ms07} and \citet{bh12}, we find that the
space density of radio BCGs is significantly lower than AGNs when
$P_{\rm1.4~GHz}\gtrsim 10^{24.5}~\rm W~Hz^{-1}$ or lower than
star-forming galaxies when $P_{\rm1.4~GHz} \le 10^{24.5}~\rm
W~Hz^{-1}$. The slope and the tuning points are very different from
those of radio luminosity functions of AGNs and star-forming galaxies.

\begin{table}
\setlength{\tabcolsep}{1.5mm}
{\footnotesize
\begin{center}
  \caption{Fitting parameters of radio luminosity functions of BCG samples.}
  \label{tab5}
\begin{tabular}{lcrrl}
\hline
\hline
\mc{1}{l}{BCG Sample} &\mc{1}{c}{$C_{0}$} &\mc{1}{c}{$P_{0}$} &\mc{1}{c}{$\alpha$} &\mc{1}{c}{$\beta$}\\
\hline
All                            & 9.4$\pm$1.4  &39.9$\pm$5.5 &3.43$\pm$0.79 &0.12$\pm$0.06       \\
$M\rm_{r}^{e}\ge-23.3$          & 6.3$\pm$0.6  &32.4$\pm$2.4 &4.46$\pm$0.81 &0.08$\pm$0.04       \\
$M\rm_{r}^{e}<-23.3$            & 4.5$\pm$1.1  &50.9$\pm$11.1 &3.37$\pm$1.24 &0.14$\pm$0.08       \\
$35<R\rm_{L*}\le45$             & 5.8$\pm$0.8  &33.6$\pm$3.8 &3.59$\pm$0.67 &0.09$\pm$0.05       \\
$R\rm_{L*}>45$                  & 5.1$\pm$1.0  &44.9$\pm$7.8 &3.48$\pm$0.97 &0.13$\pm$0.07       \\
$\Gamma\le-0.3$               & 5.5$\pm$0.7  &30.5$\pm$3.1 &3.75$\pm$0.77 &0.08$\pm$0.05       \\
$\Gamma>-0.3$                 & 5.6$\pm$1.1  &47.2$\pm$8.4 &3.40$\pm$1.00 &0.12$\pm$0.07       \\
\hline
\mc{5}{l}{Notes: $C_{0}$ in $10^{-9}\rm~Mpc^{-3}~dex^{-1}$, $P_{0}$ in $10^{24}\rm~W~Hz^{-1}$.}
\end{tabular}
\end{center}
}
\end{table}

To check the dependence of radio luminosity functions on the BCG and
cluster properties, we divide the 1,594 radio BCGs into two half
sub-samples according to the BCG absolute magnitude, cluster richness
and also dynamical parameter of clusters. The radio luminosity
function of every two subsamples are shown in Figure~\ref{fig9} and
the best fitting parameters are listed in Table~\ref{tab5}. The
functions are normalized according to BCG numbers for easy comparison.
We do see different radio luminosity functions for BCGs of different
ranges of absolute magnitude or different dynamical states. However,
no significant difference can be found for the two BCG samples of
slightly different richness ranges. The results imply that more radio
power BCGs are associated with optically bright BCGs in the relaxed
clusters, confirming the result in Section~\ref{rc} and conclusions by
\citet{mhr+09} and \citet{kvc+15}.

\section{Summary and conclusions}

Basing on the largest optical catalog of galaxy clusters of WH15 and
the largest radio survey database of the NVSS and FIRST, we identified
a large sample of 7,138 radio loud BCGs.

We found that radio power data of BCGs are rather scattered when they
are plotted against the BCG absolute magnitudes and cluster mass
proxies or cluster dynamical parameters. Very weak or no significant
correlations can be found between the radio power and the BCG or
cluster parameters. The fraction of radio laud BCGs has been checked
against the BCG absolute magnitudes and cluster richness, and we
confirm that radio loud fraction of BCGs does increase with BCG
luminosity and cluster richness in two-dimension. By using the large
complete BCG sample, we construct the radio luminosity functions of
BCGs, and do not find any redshift evolution in the redshift range of
$0.05<z\le0.45$. Radio luminosity functions are different for BCGs
with different ranges of BCG luminosity and the cluster dynamical
parameter. We conclude that BCGs are more probably radio powerful if
they have a larger absolute magnitude and resident in more relaxed
cluster.

\section*{Acknowledgments}
The authors are supported by the National Natural Science Foundation
(No. 11473034) and by the Strategic Priority Research Program ``The
Emergence of Cosmological Structures'' of the Chinese Academy of
Sciences, Grant No. XDB09010200.
Funding for SDSS-III has been provided by the Alfred P. Sloan
Foundation, the Participating Institutions, the National Science
Foundation, and the US Department of Energy.  The SDSS-III Web site is
http://www.sdss3.org/.

\bibliographystyle{mnras}
\bibliography{ref}
\label{lastpage}
\end{document}